\documentclass[apj,twocolappendix,appendixfloats, numberedappendix]{emulateapj}
\usepackage{bm,graphicx,url}
\usepackage{natbib}

\begin{document}

\title{The peculiar light-curve of J1415+1320: A case study in extreme scattering events}
\author{H.~K.~Vedantham$^{1}$, A.~C.~S.~Readhead$^1$, T.~Hovatta$^{2,3,4}$, L.~V.~E.~Koopmans$^5$, T.~J.~Pearson$^1$, R.~D.~Blandford$^6$, M.~A.~Gurwell$^7$ , A.~L\"ahteenm\"aki$^{2,3}$, W.~Max-Moerbeck$^8$, V.~Pavlidou$^9$, V.~Ravi$^{1}$, R.~A.~Reeves$^{10}$, J.~L.~Richards$^{1}$, M.~Tornikoski$^2$, J.~A.~Zensus$^8$}
\affiliation{$^1$ Owens Valley Radio Observatory, California Institute of Technology,  Pasadena, CA 91125, USA} 
\affiliation{$^2$ Aalto University Mets\"ahovi Radio Observatory,  Mets\"ahovintie 114, 02540 Kylm\"al\"a, Finland} 
\affiliation{$^3$ Aalto University Department of Radio Science and Engineering, Finland} 
\affiliation{$^4$ Tuorla Observatory, Department of Physics and Astronomy,  University of Turku, Finland} 
\affiliation{$^5$ Kapteyn Astronomical Institute, University of Groningen, PO Box 800, NL-9700 AV Groningen, the Netherlands}
\affiliation{$^6$ Kavli Institute for Particle Astrophysics and Cosmology, Department of Physics, and  SLAC National}
\affiliation{Accelerator Laboratory, Stanford University, Stanford, CA 94305, USA} 
\affiliation{$^7$ Harvard-Smithsonian Center for Astrophysics, Cambridge, MA 02138, USA} 
\affiliation{$^8$ Max-Planck-Institut fur Radioastronomie, Auf dem H\"ugel 69, D-53121 Bonn, Germany} 
\affiliation{$^9$ Department of Physics and Institute of Theoretical and Computational Physics, University of Crete, 71003 Heraklion, Greece, and Foundation for Research and Technology -- Hellas, IESL, 7110 Heraklion, Greece} 
\affiliation{$^{10}$ CePIA, Astronomy Department, Universidad de Concepci\'on,  Casilla 160-C, Concepci\'on, Chile} 


\begin{abstract}
The radio light-curve of J1415+1320 (PKS 1413+135) shows time-symmetric and recurring U-shaped features across the cm-wave and mm-wave bands. The symmetry of these features points to lensing by an intervening object as the cause. U-shaped events in radio light curves in the cm-wave band have previously been attributed to Extreme scattering events (ESE). ESEs are thought to be the result of lensing by compact plasma structures in the Galactic interstellar medium, but the precise nature of these plasma structures remains unknown. Since the strength of a plasma lens evolves with wavelength $\lambda$ as $\lambda^2$, the presence of correlated variations at over a wide wavelength range casts doubt on the canonical ESE interpretation for J1415+1320. In this paper, we critically examine the evidence for plasma lensing in J1415+1320. We compute limits on the lensing strength, and the associated free-free opacity of the putative plasma lenses. We compare the observed and model ESE light curves, and also derive a lower limit on the lens distance based on the effects of parallax due to the Earth's orbit around the Sun. We conclude that plasma lensing is not a viable interpretation for J1415+1320's light curves and that symmetric U-shaped features in the radio light curves of extragalactic sources do not present {\em prima facie} evidence for ESEs. The methodology presented here is generic enough to be applicable to any plasma lensing candidate. 
\end{abstract}
\maketitle
%
%
%
%
%
%
%
%
\section{Introduction}
The 40\,m telescope at the Owens Valley Radio Observatory (OVRO) has been monitoring about 1800 blazars at 15\,GHz, with a cadence of twice a week, since 2008, in support of {\em Fermi-GST} \citep{ovro40m}. The OVRO sample consists of $\gtrsim 0.1$\,Jy flat spectrum radio quasars (FSRQ). The radio emission from FSRQs in the cm-wave band is expected to be dominated by a compact core. The compactness of sources in the OVRO sample makes it an excellent resource to study propagation effects such as lensing. 

Lensing is a result of ray deflection caused by spatial fluctuations in the refractive index of the medium of propagation. In the context of interstellar ray propagation, electron density fluctuations in interstellar plasma provide the refractive index variation. The natural length scale probed by this effect is given by the Fresnel scale, which for Galactic plasma refraction is of the order of $\sim 0.1$\,AU. The study of plasma refraction is therefore a unique probe of the interstellar medium on such small spatial scales.

\begin{figure*}
\centering
\includegraphics[width=\linewidth]{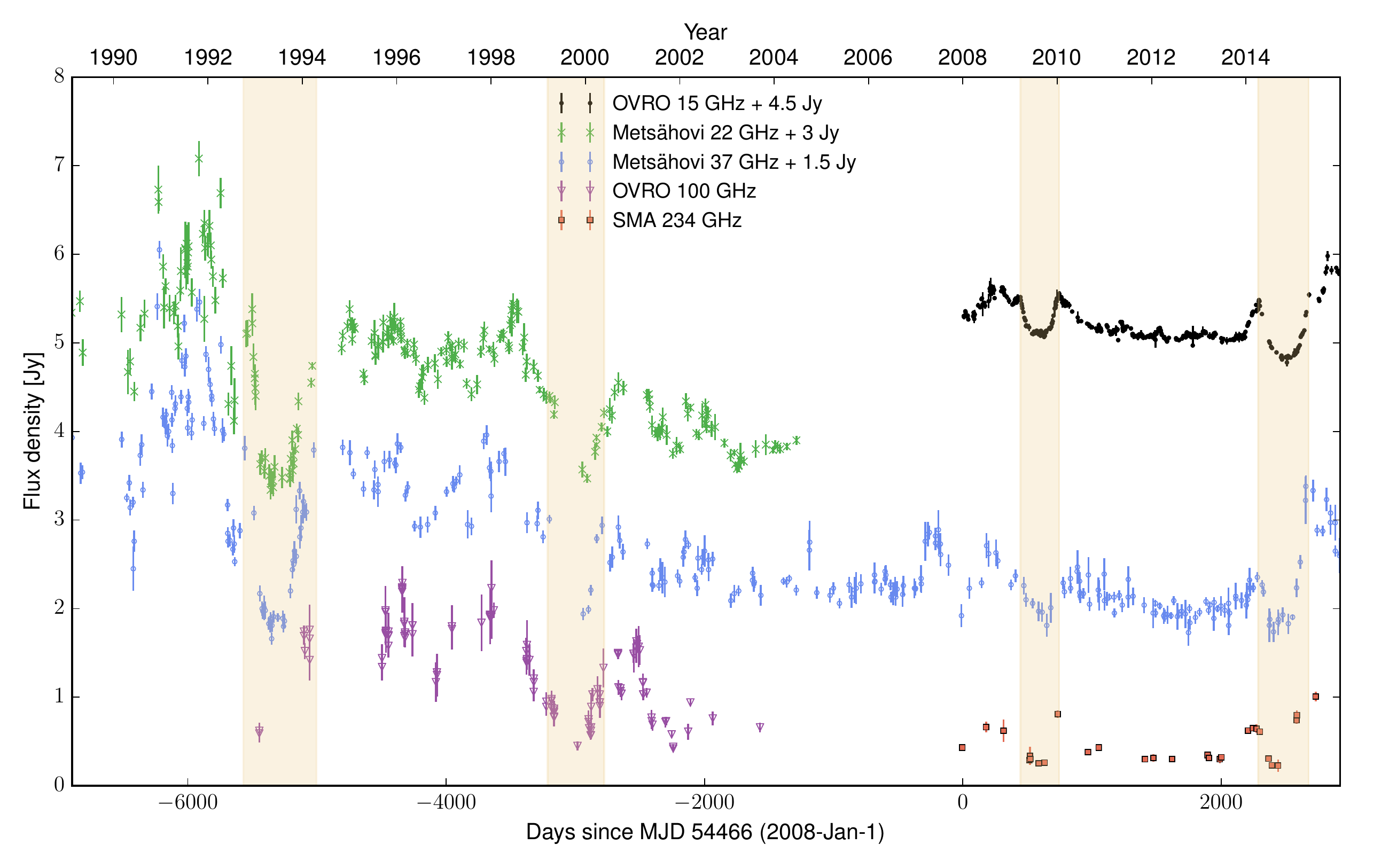}
\caption{Multi-frequency multi-decade radio light curve of J1415+1320 from various monitoring programs. Flux density offsets have been added to the light curves for clarity. The peculiar U-shaped events have been highlighted.\label{fig:light_curves}}
\end{figure*}

Lensing typically leads to multiple (de-)magnified images of the lensed source. Even if individual images cannot be resolved, flux-density variations due to the relative motion in the source--lens--observer system may be observed in light curves. The OVRO survey with its low, arcmin-scale angular resolution falls into this category. Separating lensing induced variability from intrinsic variability is difficult. Lensing however must, on average, yield time-symmetric features. We undertook a manual search for such symmetric events in the OVRO archive. Our search revealed several lensing candidates, but one source, J1415+1320 (PKS 1413+135), stood out for several reasons. J1415+1320 shows recurring, chromatic and highly symmetric U-shaped features in its light curve which are unlike anything reported so far (see Fig. \ref{fig:light_curves} and \S 1.2 below). A collation of multi-frequency data from the Mets\"ahovi radio observatory, the Sub-Millimeter Array (SMA), and OVRO shows that the U-shaped events persist even in the mm-wave band.
%
\subsection{The mystery of U-shaped events}
%
U-shaped events in the light curves of compact extragalactic sources were first reported by \citet{fiedler1987}, who named them extreme scattering events (ESE). The high degree of symmetry and chromaticity (frequency dependence) of such events led \citet{fiedler1987} to interpret them as lensing by intervening plasma structures. The plasma lenses were inferred to possess high column densities ($N_e\sim 10^{17}$\,cm$^{-2}$) over relatively small transverse extents ( $\lesssim 10$\,AU). Such lenses will be highly over-pressurized (density of $n_e\sim 10^3$\,cm$^{-3}$), and therefore cannot exist in pressure balance with the ambient interstellar medium ($n_e\sim 0.03$\,cm$^{-3}$). To mitigate this problem, two classes of models were put forth. In the first class of models, the lenses are ionized sheaths of self-gravitating sub-stellar objects \citep{esedm}. Such objects will however have to form a considerable fraction of Galactic dark matter if they are to generate the observed occultation rates \citep{esedm}. The second class of models invokes highly elongated plasma sheets seen edge-on, such that the required column density can be obtained at modest over-densities \citep{gssheet,pksheet}. Whether plasma sheets with such large axial ratios indeed exist remains unknown, and ESEs have remained an unsolved mystery for over 3 decades.
%
\subsection{The peculiar light-curve of J1415+1320}
%
The implied ESE rate based on previous surveys is about $7\times 10^{-3}$\,source$^{-1}$\,year$^{-1}$ \citep{fiedler1987}. Based on this, J1415+1320 should on average undergo an ESE every $140$\,years. We instead observe four ESE-like events in about 25\,years. The strength of a plasma lens evolves with wavelength as $\lambda^2$, and ESEs have so far only been seen at $\lambda_{\rm cm}\gtrsim 2$ \citep{pushkarev2013,fiedler1987, bannister2016}. J1415+1320 instead shows correlated ESE-like variability down to $\lambda_{\rm cm}=0.1$. First, this implies a particularly strong and compact plasma lens, further compounding the ESE mystery. Next, the observed magnitude of (de-)magnification is nearly achromatic over a factor of 20 in wavelength, which brings into question the very hypothesis of plasma lensing. These inconsistencies prompted us to take a fresh look at the viability and consequences of the plasma lensing hypothesis--- the principal aim of this paper.
%

%
%
\subsection{Outline of the paper}
To place the peculiar light curves of J1415+1320 in context, we describe the properties of the source in \S 2. We then discuss the feasibility and implications of plasma lensing in \S 3. We end with a discussion of our main findings in \S 4. We note that though \S 3  is geared towards explaining the peculiar light-curve of J1415+1320, most aspects of our analysis are generic enough to be applicable to any plasma-lensing candidate in the radio band. As such, this section is a case study in plasma lensing of extragalactic radio sources. 
%
%
%
%
%
%
\section{Source properties and lensing geometry}
We now briefly summarize the properties of J1415+1320, and the plausible lensing geometries.
\subsection{Optical association}
J1415+1320 is classified as a blazar (on-axis AGN) of the BL-Lac type \citep{beichman1981}. The core of the radio source is co-incident with the optical light-centroid of an edge-on spiral galaxy at $z\approx 0.24$ \citep[][HI absorption data]{carilli1992}, albeit an angular offset of 13\,mas (50\,pc) has been reported based on isophotal analysis of \emph{Hubble} images \citep{perlman2002}. In addition, bright radio sources such as J1415+1320 are almost exclusively associated with giant elliptical galaxies. These factors have led to suggestions that the radio source is a background object unrelated to the spiral \citep{stocke1992, perlman1994}. If so, then the absence of multiple images due to lensing by the bulge of the spiral, constrains the radio source to be no further than a redshift of $z\lesssim 0.5$ \citep{lamer1999}.

Optical spectroscopy of the spiral galaxy by \citep{vedantham_sci} shows several lines consistent with a redshift of $z=0.247$ as expected. There is no evidence for emission from another redshift. No broad lines from the AGN were detected, most likely due to extinction caused by dust-lanes in the spiral. Finally, no excess line emission around 6563$\pm50$ Angstrom (Galactic H$\alpha$), is seen up to a 3$\sigma$ limit of 2.5\,mJy.


\subsection{Parsec-scale radio structure}
VLBI imaging of J1415+1320 reveals a relatively young radio source with a bright core and a two sided jet-lobe structure straddling the core \citep{perlman1996}. Based on this morphology, J1415+1320 belongs to the family of compact symmetric objects (CSO). The bulk of the observed radio flux shortward of about 2\,cm comes from the bright unresolved core \citep{perlman1996}. For a jet Lorentz factor of  $\gamma=1$, a flux-density of $1$\,Jy, and an equipartition brightness temperature of $T_b=5\times 10^{10}$\,K, the expected angular size of the core is $\sigma_s(\lambda) = 70\lambda_{\rm cm}\,\mu$as. Here the source is assumed to be a Gaussian with FWHM of $2.355\sigma_s$. The core remains unresolved in 15 and 43\,GHz VLBI images \citep{perlman2002}, which independently yield $\sigma_s (15\,{\rm GHz})<200\,\mu$as and $\sigma_s(43\,{\rm GHz})<85\,\mu$as, respectively. The core however may internally consist of compact $\mu$as scale components whose brightness temperature may be as high as $T_{\rm b}\sim 10^{13}$\,K as has been observed in other sources \citep{johnson2016}.

\subsection{Angular broadening}
Angular broadening due to scattering in interstellar plasma will have an additional influence on the apparent size of the source. Since the source is at a high Galactic latitude ($b\approx 65$\,deg) angular broadening in the Milky way is expected to be very small \citep[$\sigma_{\rm sc}\lesssim 1\,\mu$as at 15\,GHz;][]{ne2001}. Angular broadening in the interstellar medium of the putative spiral host however, could be significantly larger due to its edge-on configuration, and because the radio source is seen through a giant molecular cloud in the spiral \citep{gmc}. Very little is known about the turbulent properties of interstellar plasma in other galaxies. We may however obtain rough estimates by assuming that the spiral has a `Milky Way like' interstellar medium, and by appropriately scaling the value of angular broadening expected along an equivalent sight-line in the Milky-way. To account for the angular offset between the radio source and the optical light-centroid of the spiral, we consider a Galactic longitude and latitude of 0\,deg and 0.5\,deg respectively. The expected angular broadening for such a Galactic sight-line is about 0.5\,mas at 15\,GHz \citep{ne2001}. In addition, the magnitude of angular broadening must be scaled by the geometric `lever-arm' as\footnote{We use the standard notation wherein $D_l$, $D_s$ and $D_{ls}$ are the observer--lens, observer--source and lens--source angular diameter distances respectively.} $D_{ls}/D_s$. If the radio source is in the spiral, the lever-arm is small and of the order $D_{ls}/D_s\sim 10\,{\rm kpc}/824\,{\rm Mpc}\approx 10^{-5}$.  The expected angular broadening in this geometry is at a sub-$\mu$as level and can be neglected. If instead, the radio source is behind the spiral, for $D_{ls}=300$\,Mpc, the expected angular broadening is $0.5\,{\rm mas}\times D_{ls}/D_s \approx 180\,\mu$as. This value is marginally inconsistent with multi-frequency VLBI maps of J141+1320 \citep{perlman1996}. We note however that the interstellar medium is known to be `patchy' and the line of sigh to J1415+1320 may pass through a region of significantly lower turbulence. Hence, the above cannot be admitted as evidence against the background-source hypothesis.
\subsection{Plausible lensing geometries}
\begin{figure*}
\centering
\includegraphics[width=\linewidth]{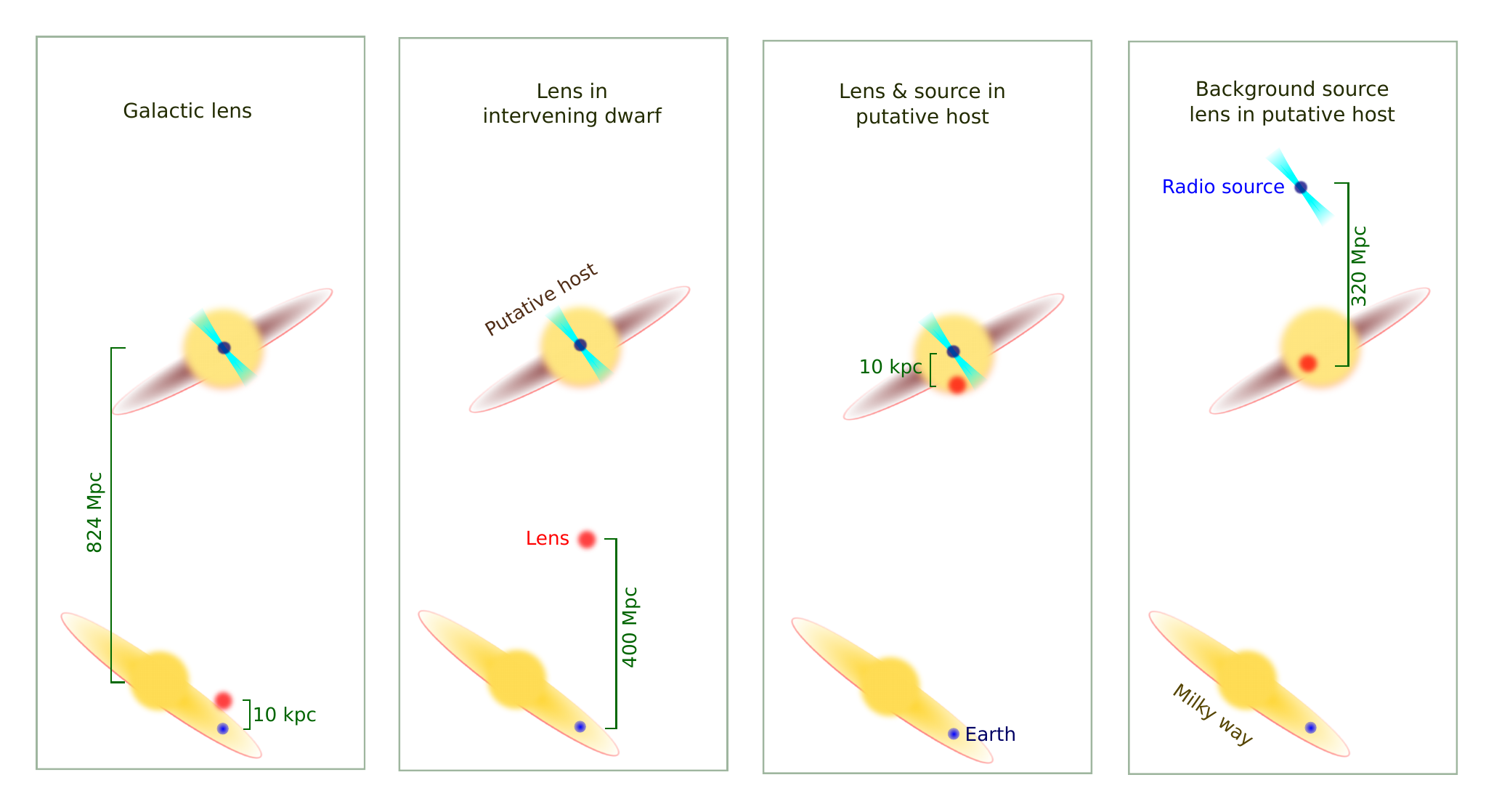}
\caption{Cartoon showing the four lensing geometries under consideration and typical distances in the observer--lens--source system.\label{fig:lensing_geometry}}
\end{figure*}

Given the above known properties of J1415+1320, we consider 4 plausible lensing geometries arranged in increasing distance between the Earth and the lens (see Fig. \ref{fig:lensing_geometry} for a cartoon depiction). 
\begin{itemize}
\item The lens is a Galactic object. Since $D_s\gg D_l$, whether the radio source is associated with the spiral or not is inconsequential. The high galactic latitude of J1415+1320 ($b\approx 65$\,deg) implies that the lens distance is likely not larger than the scale height of the Galactic disc: $D_l\lesssim 0.3$\,kpc, failing which the lenses will have to be an exotic halo population. 

\item The lens is in some intervening system unrelated to both the spiral and the Milky way. Here the association of the radio source with the spiral has only a moderate effect on lensing parameters. Since no intervening galaxy has been detected, this model essentially requires the lens to be associated with an optically faint object such as a distant isolated globular cluster or a faint dwarf galaxy.

\item Both the source and lens are in the spiral galaxy at $z=0.24$. This gives, $D_l\approx D_s = 824$\,Mpc, and due to the edge-on configuration of the spiral, we assume $D_{ls}\lesssim 30$\,kpc.

\item The lens resides in the spiral ($D_l=824$\,Mpc), but the radio source is an unrelated background object. As argued in \S 2.1, $z\lesssim 0.5$, which gives $D_{ls}\lesssim 320$\,Mpc.
\end{itemize}
We now explore the feasibility of, and evidence for the plasma lensing hypothesis for the above geometries.
%
%
%
%
%
\section{Plasma lensing}
\begin{figure}
\centering
\includegraphics[width=\linewidth]{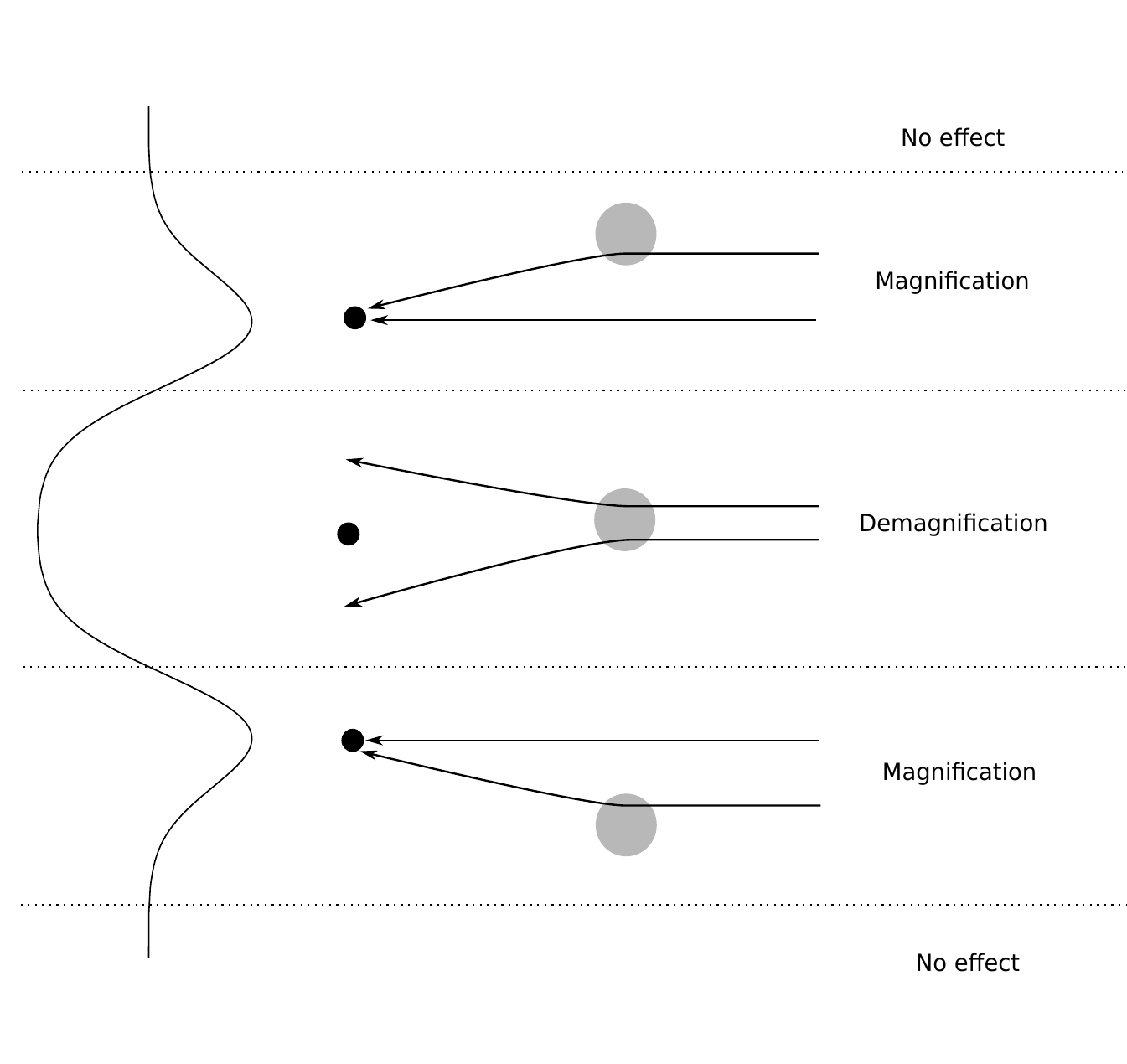}
\caption{A cartoon depiction (not to scale) of the effect of a diverging plasma lens on a background source. Black circle is the Earth, and the gray circle is the lens. Relative motion in the source--lens--observer system leads to symmetric magnification events that bracket a de-magnification trough as shown in the U-shaped curve on the left.\label{fig:ese_cartoon}}
\end{figure}

The refractive index of plasma is less than unity. A plasma lens therefore diverges light rays. This leads to de-magnification of occulted sources when the observer--lens--source all lie on a straight line. At small displacements, curvature in the lens-profile leads to magnification events that bracket the de-magnification trough. This leads to the characteristic U-shaped profile seen in extreme scattering events. A cartoon depiction of this phenomenon is shown in Fig. \ref{fig:ese_cartoon}. We now determine the minimum lensing strength required to explain the light curve of J1415+1320.

%
%
\subsection{Minimum lensing strength}
\begin{figure}
\includegraphics[width=\linewidth]{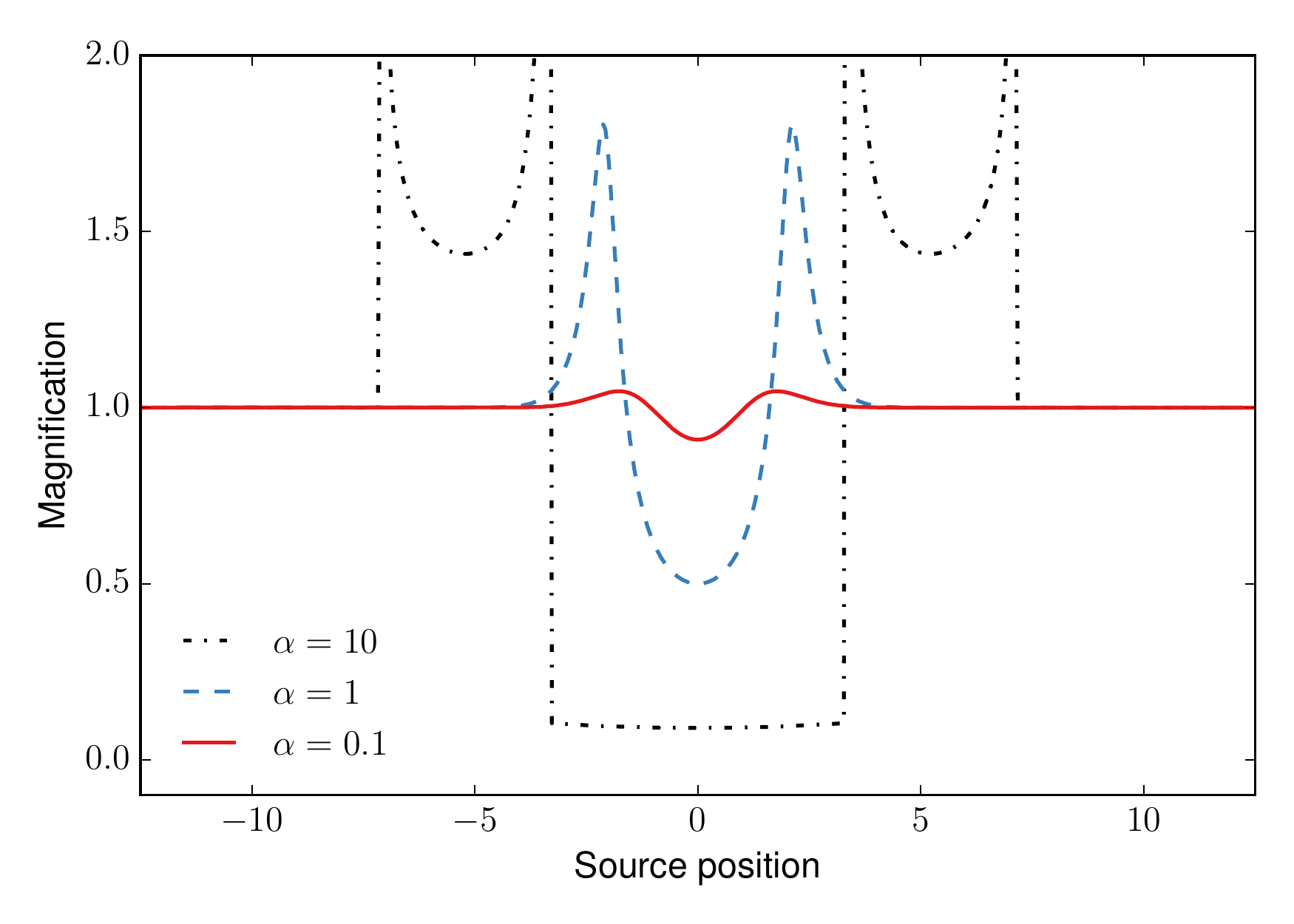}
\caption{Light curves for a point-like source occulted by gaussian plasma lens of unit size ($\sigma_l=1$). \label{fig:ldef} }
\end{figure}

%
%
The strength of a lens depends on its intrinsic properties as well as the lensing geometry. These two properties can be combined into a dimensionless `lensing potential' $\Psi(\bm{\theta})$ where $\bm{\theta}$ is the 2-dimensional image position. The source position $\bm{\beta}$ corresponding to image position $\bm{\theta}$ is then given by the lensing equation: $\bm{\beta} = \bm{\theta} - \nabla_{\bm{\theta}} \Psi(\bm{\theta})$.  For a given $\bm{\beta}$, the lens equation may have more than one solution in $\bm{\theta}$--- a case that corresponds to multiple images of the background source. \\
The lensing potential for a generic plasma lens is given by \citep{tuntsov2016}
\begin{equation}
\label{eqn:ese_pot}
\Psi(\bm{\theta}) \propto \frac{D_{ls}}{D_lD_s}r_e\lambda^2N_e(\bm{\theta})
\end{equation}
where $r_e$ is the classical electron radius and $N_e(\bm{\theta})$ is the electron column density within the lens. Lensing requires two conditions to be satisfied: (i) the source angular size should be comparable to or less than the lens angular size, and (ii)  the lens should possess sufficient strength to deflect light-rays by angles comparable to the angular size of the lens. Consider for instance, a Gaussian lens of the form
\begin{equation}
\label{eqn:glens}
\psi(\bm{\theta}) = \alpha(\lambda) \exp\left( -\frac{\left|\bm{\theta}\right|^2}{2\sigma_l^2}\right).
\end{equation}
The lens has an effective solid angle of $\Omega_l = 2\pi\sigma_l^2$, and the strength parameter
\begin{equation}
\label{eqn:gauss_pot}
\alpha(\lambda) = \frac{1}{2\pi\sigma_l^2}\frac{D_{ls}}{D_lD_s}N_er_e\lambda^2,
\end{equation}
where $N_e=N_e(\bm{0})$ is the transverse gradient of the lensing potential on angular scales comparable to the size of the lens itself. If $\alpha\ll 1$, the effect of the lens is negligible. For $\alpha \lesssim 1$, (de-)magnification is moderate, and the lens-equation has a single valued solution (no multiple images). $\alpha\gtrsim 1$ leads to multiple images, and large (de-)magnification. 
Fig. \ref{fig:ldef} shows the light curves for a point-like source occulted by a plasma lens with a column density profile given by equation \ref{eqn:glens} for $\alpha = 0.1$, 1 and 10. For $\alpha\gtrsim 1$, 2 pairs of caustic are visible. We have simulated several other (non-gaussian) lens profiles, and find that while the inner pair of caustics are ubiquitous, the presence of an outer caustic pair is not a generic property.

J1415+1320 shows large variations in magnification over a factor of 20 in wavelength, or a factor of 400 in lensing potential. 
A lower bound on the lensing potential is set by the requirement for lensing to have a significant effect on ray propagation at the shortest wavelength of $\lambda_{\rm cm}=0.1$. More formally, we require $\alpha(\lambda = 0.1{\rm \,cm})>1$. Normalizing to a nominal lens size\footnote{This will make the lens comparable in size to the upper-limits on the source at 15\,GHz.} of $\sigma_l=100\,\mu$as, equation \ref{eqn:gauss_pot} yields the constraint
\begin{equation}
\label{eqn:min_lens_str}
\frac{D_{ls}}{D_lD_s}N_e > 5\times10^{-4}\,\left(\frac{\sigma_l}{100\,\mu{\rm as}}\right)^2\,\,\,\,{\rm cm}^{-3}
\end{equation}
Equation \ref{eqn:min_lens_str} assumes a point-like source. The angular size of the source relative to that of the lens, $\sigma_s$, and the lensing strength, $\alpha$ may be simultaneously increased to generate similar light curves (see Fig.\ref{fig:amb} for a demonstration). This degeneracy between $\alpha$ and $\sigma_s$ implies that equation \ref{eqn:min_lens_str} must be treated as a conservative lower limit.

We now explore the consequences of the implied column density from equation \ref{eqn:min_lens_str}. We discuss the consistency of the observed light curves with a Gaussian lens model constrained by equation \ref{eqn:min_lens_str} in \S 3.4.
%
%
\subsection{Free-free opacity}
Table \ref{tab:plasma} summarizes the direct consequence of the limits set in equation \ref{eqn:min_lens_str}. We have assumed an isotropic lens---the lens has the same line-of-sight size as its transverse size, and obtained the electron density, $n_{\rm e}$, which in turn yields the free-free optical depth, $\tau_{\rm ff} \propto \int n^2_{\rm e}\,{\rm d}l$ (details in Appendix). The free-free optical depth must be significantly smaller than unity, failing which radio photons cannot freely pass through the lens and the magnification events from plasma lensing cannot be obtained.

All geometries with an isotropic extragalactic lens may be rejected based on their high free-free opacity. Next, we consider anisotropic sheet-like geometries viewed edge-on. For such lenses, the free-free opacity from Table \ref{tab:plasma} must be reduced by a factor equal to the axial ratio: $\zeta = $depth/width of the lens. All extragalactic lenses still require unrealistic axial ratios of $\gtrsim 10^3$ to allow passage of 15\,GHz photons. 

The free-free opacity value scales as $\sigma_l^3$, and it is therefore appealing to significantly reduce the lens size to $\lesssim 10\,\mu$as by invoking lensing of compact $\sim \mu$as scale source components. For all extragalactic lens models, this scenario is problematic for the following reason. Lens modeling requires occultation of a source component of about $600$\,mJy in addition to an unlensed component (details in \S 3.4). For a relatively high brightness temperature of $T_{\rm b}=10^{13}$\,K, the implied component size is of order $\sigma_s\sim 10\,\mu$as, which could be lensed by a $\sim 10\,\mu$as scale lens. A relativistically moving 600\,mJy source component can cross four $10\,\mu$as-scale extragalactic lenses separated by $\sim 50\,\mu$as, to yield four year-long events separated by five years each. The distance traversed by such a source component is about $200\,\mu$as with respect to a presumably unlensed stationary core of comparable brightness. Source components of such high flux-densities have not been detected in sub-mas resolution imaging of J1415+1320 \citep{perlman2002,mojave}. Scenarios where the four U-shaped events are caused by four different 600\,mJy source components crossing the same lens are also problematic since they require a contrived arrangement where the source components remain relatively stable during the ESE and `cool' rapidly thereafter to maintain the total flux density of the source at the $\sim 1$\,Jy level as observed. Hence, apart from free-free opacity constraints, extragalactic lens models have an additional problem in explaining the timing, duration, and flux-densities of the four ESE-like events. We therefore find Galactic lenses at $D_l=10$\,kpc (value justified in \S 3.3) to be the most viable ESE scenario among the four geometries.
\begin{table*}
\caption{Table showing lower bounds on plasma density and free-free opacity for the putative plasma lenses. Rows correspond to different lensing geometries (see Fig. \ref{fig:lensing_geometry}). A lens size of $\sigma_l=100\,\mu$as has been assumed. Free-free opacity is proportional to $\sigma_l^3$.  \label{tab:plasma}}
\centering
\begin{tabular}{lllllll}
Model & $D_{ls}$ & $D_l$ &  $N_e$ & $n_e$ & Lens size 	& Free-free optical depth (15\,GHz)\\ \hline \\
Galactic lens 								& 800\,Mpc 	& 10\,kpc   & $10^{19}$ 	& $5\times 10^5$ 	& 1\,AU 	& $3\times 10^{-3}$\\
Lens in intervening dwarf 					& 700\,Mpc	& 400\,Mpc  & $10^{23}$		& $9\times 10^{5}$ 	& 0.2\,pc 			& $5\times 10^2$ \\
Lens and source in putative host 			& 30\,kpc	& 800\,Mpc	& $10^{29}$		& $1.1\times 10^{10}$ 		& 0.4\,pc 			& $1.7\times 10^{11}$ \\
Background source; lens in putative host	& 400\,Mpc	& 800\,Mpc	& $10^{25}$		& $1.4\times 10^{6}$ 	& 0.5\,pc 				& $2\times 10^3$\\
\\ \hline
\end{tabular}
\end{table*}
%

%
%
\subsection{Retrograde motion of galactic lenses}
\begin{figure*}
\centering
\includegraphics[width=\linewidth]{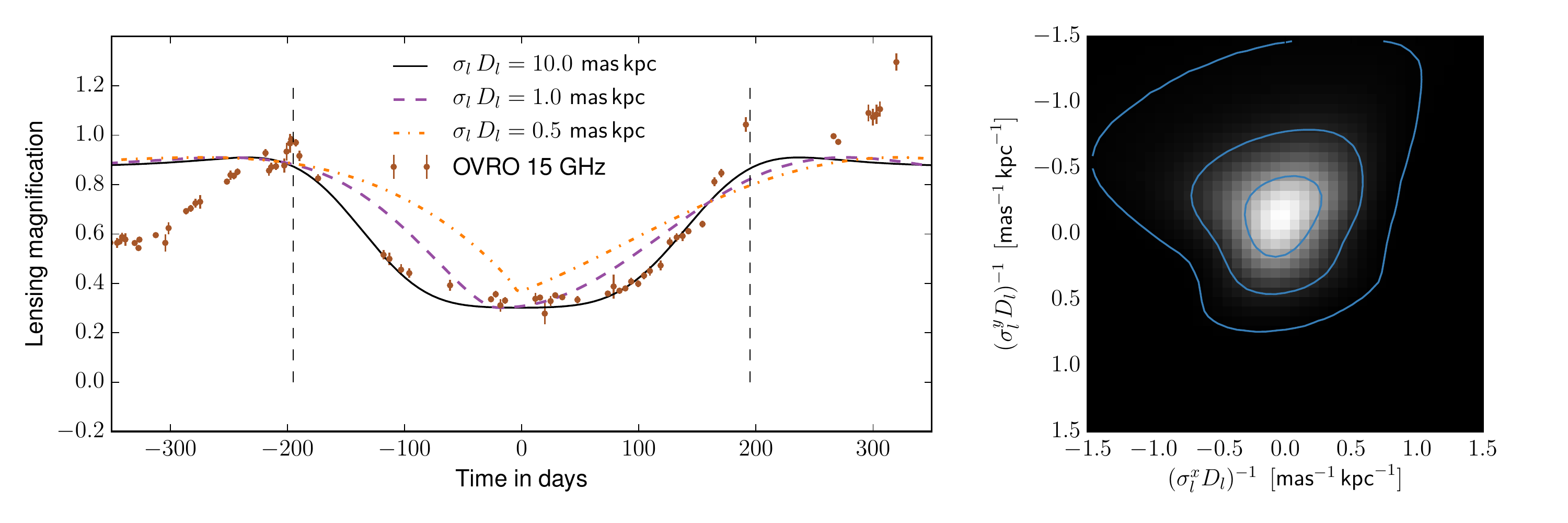} 
\caption{Effect of the annual parallax on lensing light curves for a Galactic plasma lens. Left panel shows the data (2015 event) and model light curves for varying magnitudes of this parallactic effect. Right panel shows the likelihood function (1,2, and 3\,$\sigma$ contours over-plotted) for varying magnitude of the parallactic effect. The x and y axes lie along the ecliptic plane with x-axis along zero ecliptic longitude.\label{fig:parallax}}
\end{figure*}

%
%
The symmetry of U-shaped ESEs is a result of rectilinear relative motion between the source and an axially symmetric lens. The orbit of the Earth around the Sun can perturb an otherwise rectilinear trajectory leading to asymmetries in the observed light curves. The ESE-like events presented here last about $400$\,days (interval between magnification events), which is significantly longer than previously reported ESEs. If the lenses are Galactic in origin (as implied by \S 3.2), this allows us to place unique constraints on the size of the lens as follows. Consider a plasma lens at 1\,kpc from the Earth. Because the parallax due to Earth's annual motion around the Sun is 1\,mas at 1\,kpc, a lens that is smaller than 1\,mas must yield a light-curve with significant asymmetry. The magnitude of this effect is inversely proportional to the product of the lens angular size and distance, $\sigma_lD_l$. To test for the presence of this `annual parallax' in the data, we simulated light curves for various values of $\sigma_lD_l$ and evaluated the residues between the data and our model within the U-shaped trough at 15\,GHz. We assigned the lowest least square error value thus obtained as the `noise' in our data and computed the likelihood of all $\sigma_lD_l$ choices by assuming Gaussian noise statistics. Anticipating a slight degeneracy between $\sigma_lD_l$ and the exact epoch of occultation, we marginalized the likelihood over the latter. The resulting likelihood function along with the 1,2 and 3$\sigma$ contours is shown in Fig. \ref{fig:parallax} (right panel).

Using Fig. \ref{fig:parallax}, the product of lens distance and lens size may be constrained to be $\sigma_lD_l>2.0$\,mas\,kpc. $D_l$ is unknown, but the lensing scenario requires $\sigma_l\approx \sigma_s$ at 15\,GHz (see \S 3.4). The best observations constraint yields $\sigma_s<200\,\mu$as (see \S 2.2). The corresponding constraint on the lens distance therefore is $D_l>10$\,kpc. Since the source is at a high Galactic latitude ($b\approx65$\,deg), these values are clearly incompatible with the plasma lens being in the Galactic disc where most of the interstellar plasma lies. 
\subsection{Inconsistency of light curves with plasma lensing}
\begin{figure*}
\centering
\includegraphics[width=\linewidth]{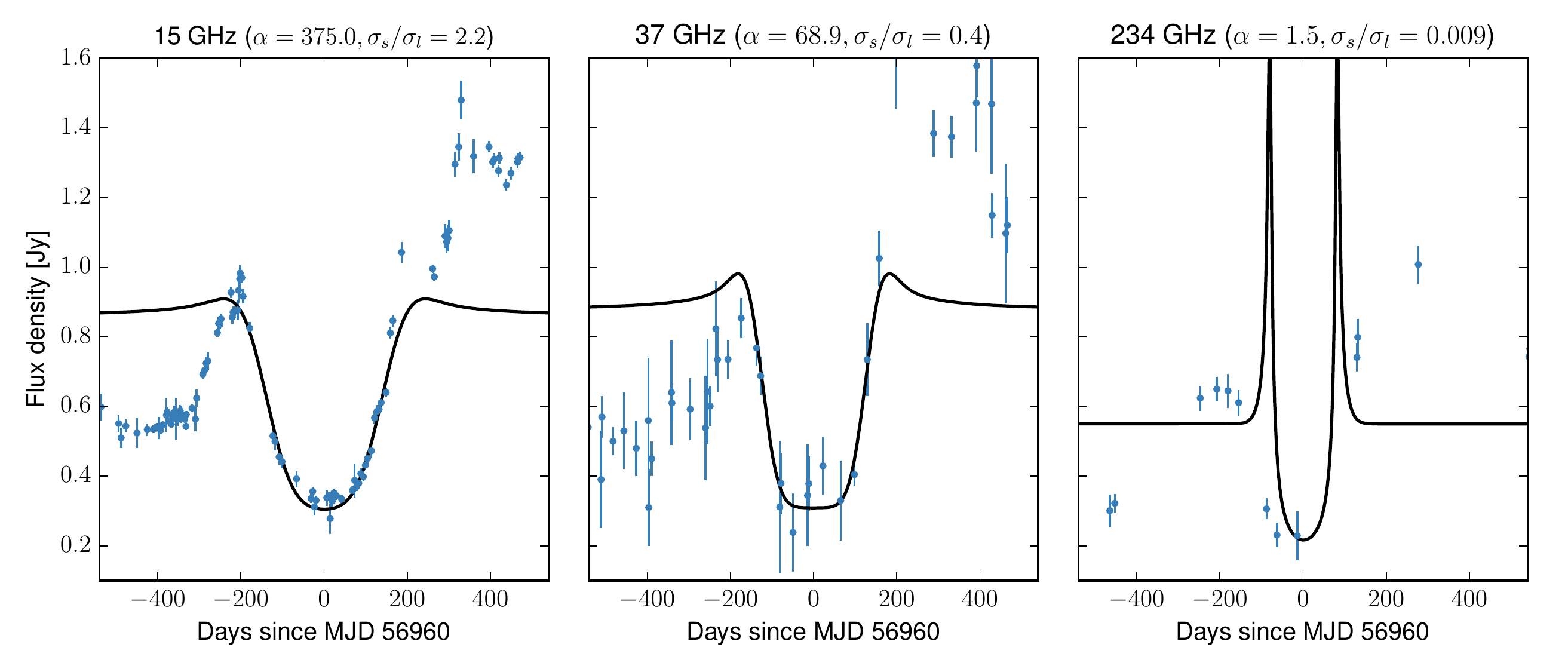}\\
\includegraphics[width=\linewidth]{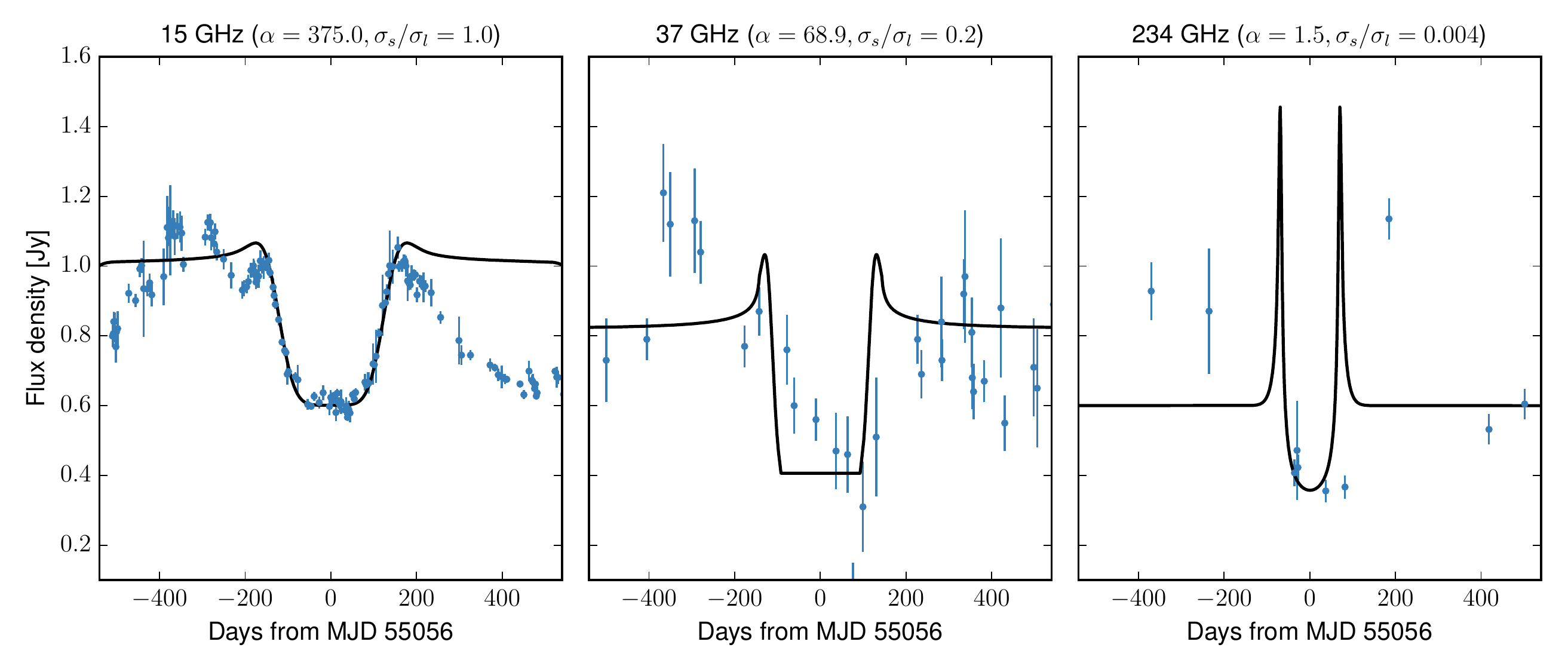}
\caption{Data overlaid with model plasma lensing light curves (Gaussian lens) for the 2010 and 2015 U-shaped events (top and bottom rows respectively). The column are for 3 frequency bands: 15, 37, and 234\,GHz spanning a factor of 400 in lensing strength, $\alpha$. The ratio between source size and lens size, $\sigma_l/\sigma_s$ is given in the panel labels. \label{fig:p1_gauss}}
\end{figure*}

Fitting of model ESE light-curves to data is complicated by the fact that almost identical light curves can be obtained by increasing the lensing strength while appropriately increasing the source size (see Fig. \ref{fig:amb} for a demonstration). We therefore proceed with the most conservative assumption: the source is point-like at the highest frequency, 234\,GHz. We thus obtain the smallest value of $\alpha(234\,{\rm GHz})$ that can account for the range of variability seen in the data. This value of $\alpha$ can then be scaled to lower frequencies, and the source size adjusted to obtain reasonable fits to the data. We found that matching the models to the data required (i) a source size that scales with wavelength as $\lambda^2$ ($\sigma_s/\sigma_l\gtrsim 1$ is required at 15\,GHz), and (ii) a 600\,mJy component of the source was lensed at 37 and 15\,GHz. A significantly smaller flux-density will not yield the required depth in the U-shaped trough. If on the other hand, the entire source is lensed, the large demagnification at 15\,GHz should have made the source virtually undetectable at 15\,GHz for about a year. The resulting model light curves are plotted with the data at 15, 37, and 234\,GHz in Fig. \ref{fig:p1_gauss} for the 2010 and 2015 events (upper and lower row respectively). 

The model provides poor fits to the data outside the de-magnification trough. Part of the reason for this may be our pre-supposed (and hence arguably inaccurate) lens profile, and intrinsic source variability. But the main reason is the generic incompatibility of the data with a de-magnification trough that is a hallmark of plasma lensing. Consider for instance, the 15\,GHz flux-densities of the source (see Fig. \ref{fig:light_curves}) at years 2009.5 and 2011.5. A de-magnification should have yielded significantly lower flux in 2009.5 as compared to 2011.5. The data instead show the source to be of comparable brightness at the two epochs.

\section{Discussion and summary}
J1415+1320 shows recurring and symmetric U-shaped features in its 3\,decade long light curves at frequencies ranging from the cm-wave band to the mm-wave band. The canonical interpretation for such U-shaped events in radio source is lensing by compact interstellar plasma structures. We have critically examined the plasma lensing interpretation and arrived at the following conclusions. 
\begin{itemize}

\item All extragalactic plasma lenses require unrealistically high axial-ratios (width/depth) of $\gtrsim 10^3$ to yield acceptable amounts of free-free opacity (Table \ref{tab:plasma}).

\item The absence of significant asymmetry expected due to the Earth's orbit around the Sun, yields a lens distance that is well beyond the Galactic disc. This places the putative Galactic lenses in a region which is expected to be largely devoid of dense interstellar material (Fig. \ref{fig:parallax}).

\item The U-shaped events seen in J1415+1320 are not well described as de-magnification events. The flux-density of the source at the bottom of the U-shaped trough is comparable to that away from the U-shaped events. For this reason, model light curves provide poor fits to the data even in the immediate vicinity of the trough (Fig. \ref{fig:p1_gauss}).
\end{itemize}

The above factors compel us to reject the ESE hypothesis for J1415+1320. We therefore conclude that symmetric U-shaped radio light curves do not present {\em prima facie} evidence for plasma lensing. This raises two pertinent questions.\\

{\em What caused the U-shaped events in J1415+1320?}. The high degree of symmetric of the U-shaped events in J1415+1320 points to lensing as the cause. The only other lensing mechanism known is gravitational lensing, which is achromatic. The viability and consequences of the gravitational lensing hypothesis are discussed in a separate paper \citep{vedantham_sci}. \\

{\em How can one ascertain and constrain plasma lensing?}. The present study of J1415+1320 has shown that largely achromatic U-shaped events may be obtained despite plasma refraction being a chromatic phenomenon. This is because an increase in source size with wavelength can partly offset the increase in lensing strength with wavelength to yield a U-shaped event with comparable levels of de-magnification over a wide wavelength range (Fig. \ref{fig:amb}). It is the incompatibility of data with model ESE light-curves over a longer-term, among other constraints, that enables us to rule out the ESE hypothesis for J1415+1320. In the absence of longer-term monitoring data, the {\em cessation} of ESE-like variability at some high frequency may be taken as evidence in favor of the ESE hypothesis. 

\section*{Acknowledgements}
The authors thank Sterl Phinney and Shrinivas Kulkarni for useful discussions. The OVRO 40-m program has been supported by NASA grants NNG06GG1G,  NNX08AW31G, NNX11A043G, and NNX13AQ89G and NSF grants AST-0808050,  and AST-1109911. The Submillimeter Array is a joint project between the Smithsonian Astrophysical Observatory and the Academia Sinica Institute of Astronomy and Astrophysics and is funded by the Smithsonian Institution and the Academia Sinica. H. Vedantham is a R.~A. \& G.~B. Millikan fellow of experimental physics. T.~Hovatta was supported in part by the Academy of Finland project number 267324. R. Reeves gratefully acknowledges support from the Chilean Basal Centro de Excelencia en Astrofisica y Tecnologias Afines (CATA) grant PFB-06/2007.
\appendix
\section{{\small Lens strength--source size ambiguity}}
\begin{figure*}
\centering
\includegraphics[width=0.75\linewidth]{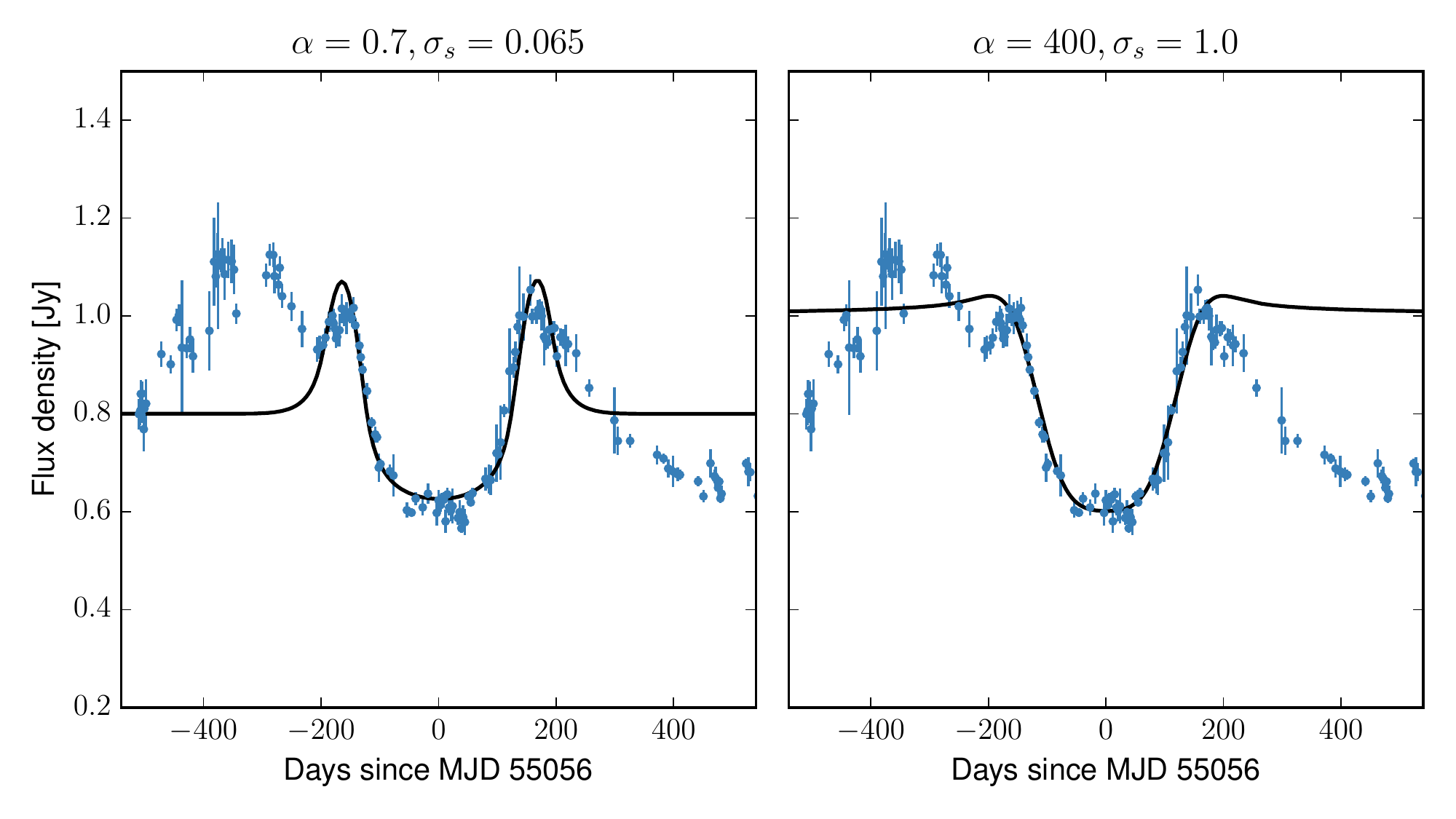}
\caption{$\alpha$--$\sigma_s$ degeneracy ($\alpha=$ lensing strength; $\sigma_s=$ source size in units of lens size. The two models shown in the two panels both describe the data-points equally well. It is therefore not possible to infer lensing parameters at a single frequency, or a closely spaced sub-set of frequencies, without explicit knowledge of the source size.\label{fig:amb}}
\end{figure*}
Many reported ESEs have either been observed at a single frequency channel, or totally uncorrelated variations are seen between 2 frequency bands. This leads to considerable problems in interpreting the light-curves. Several parameters such as lens strength, source size relative to the lens, and the flux-densities of the lensed and unlensed components must be fit to observed light curves, leading to degeneracies between parameters. In particular, we find a strong degeneracy between the source size relative to the lens and the lensing strength. As an illustration, consider the well sampled 15\,GHz ESE during 2009. In Fig. \ref{fig:amb} we have shown 2 model light curves whose lensing strengths $\alpha$, vary by a factor $>500$ and yet, the curves provide a reasonable fit to the data points. Fig. \ref{fig:amb} shows that one cannot determine the lensing strength with light curves at a single frequency. Rather only a lower limit on the lensing strength can be placed (assuming the source is significantly smaller than the lens). In our case, if only the 15\,GHz light curves were available, we would only conclude that $\alpha\gtrsim 0.7$
\section{\small Free-free opacity}
Let the transverse size of the lens be $\theta_l$, distance to the lens be $D_l$, the electron column density be $N_e$, and let the lens have a line of sight extent $\rho$ times that of its transverse extent. The electron volume density is then $n_e=N_e/(D_l\theta_l\zeta)$. The emission measure (line of sight integral of $n_e^2$) is then ${\rm EM} =  N_e^2/(D_l\theta_l\zeta)$. 

%
%

The free-free opacity at radio frequency is approximately given by
\begin{equation}
\tau_{\rm ff} = 3.28\times10^{-7}\,T_4^{-1.35}\left(\frac{\nu}{\rm GHz} \right)^{-2.1}\left( \frac{{\rm EM}}{\textrm{cm$^{-6}$\,pc}}\right),
\end{equation}

where $T=10^4T_4$ is the gas temperature in kelvin, and we assume $T_4=1$.

\end{document}